\documentclass[12pt]{article}
\usepackage[pctex32]{graphics}
\begin{document}
\vspace{2cm}
\begin{center}
{\bf  \Large  Effects of  the Shear Viscosity on the Character of Cosmological Evolution}
\vspace{1cm}

                       Wung-Hong Huang\\
                       Department of Physics\\
                       National Cheng Kung University\\
                       Tainan,70101,Taiwan\\

\end{center}
\vspace{2cm}

   Bianchi type I cosmological models are studied that contain a stiff fluid with a shear viscosity that is a power function of the energy density, such as $\zeta = \alpha \epsilon^n$. These models are analyzed by describing the cosmological evolutions as the trajectories in the phase plane of Hubble functions. The simple and exact equations that determine these flows are obtained when $n$ is an integer.   In particular, it is proved that there is no Einstein initial singularity in the models of $0\leq n < 1$. Cosmologies are found to begin with zero energy density and in the course of evolution the gravitational field will create matter. At the final stage, cosmologies are driven to the isotropic Fnedmann universe. It is also pointed out that although the anisotropy will always be smoothed out  asymptotically, there are solutions that simultaneously possess non-positive and non-negative Hubble functions for all time. This means that the cosmological dimensional reduction can work even if the matter fluid having shear viscosity. These characteristics can also be found in any-dimensional models.

\vspace{3cm}
\begin{flushleft}
E-mail:  whhwung@mail.ncku.edu.tw\\
Published in: J. Math. Phys. 31 (1990) 659-663
\end{flushleft}

%%%%%%%%%%%%%%%%%%%%%%%

\newpage
\section{Introduction}
  
  The investigation of relativistic cosmological models usually has the energy momentum tensor of matter as that due to a perfect fluid. To consider a more realistic model one may take into account the dissipative processes that are
caused by the viscosity and that have already attracted the attention of many investigators.  Misner [1] suggested that the anisotropy in an expanding universe would be smoothed out by the strong dissipative process due to the neutrino viscosity. Viscosity mechanisms in the cosmology can explain the anomalously high entropy of the present universe [2,3].  Bulk viscosity associated with the grand-unified-theory phase transition [4] may lead to an inflationary scenario [5,6].  (The inflationary cosmology invented by Guth [7] in 1981 is used to overcome several important problems arising in the standard big bang cosmology.)

    Murphy [8] obtained an exactly soluble isotropic cosmological model of the zero-curvature Friedmann model in the presence of bulk viscosity. The solutions Murphy founded exhibit the interesting feature that the big bang type singularity appears in infinite past.  Exact solutions of the isotropic homogeneous cosmology for the open, closed, and flat universe, were found by Santos et al.[9] when the bulk viscosity is the power function of energy density.  However, in some solutions, the big bang singularity of infinite density occurs at finite past. It is thus shown that, contrary to the conclusion of Murphy, the introduction of bulk viscosity cannot avoid the initial singularity in general. Anisotropic models with bulk viscosity, which is the power function of energy density, have been discussed in detail in our previous papers [10,11].

   Belinskii and Khalatnikov [12] presented a qualitative analysis about Bianchi type I cosmological models under the influence of shear viscosity.  They then found a remarkable property that near the initial singularity the gravitational field creates matter.  Recently, Banerjee et al.[13] obtained one Bianchi type I solutions for the case of stiff matter by using the assumption that shear viscosity ($\eta$) is the power  function of the energy density ($\epsilon$), i.e., $\eta= \alpha\epsilon^n$, where $\alpha$ is a constant.  However, Banerjee et al. merely analyzed the behavior of the cosmological models for some values of $n$.

       In this paper we will investigate the cosmological models again. Since we study these models by describing the evolution of cosmologies as the flow in the phase plane of Hubble functions, we can clarify the property of the models with
any value of $n$. In particular, we prove that there is no Einstein initial singularity in the models with $0\leq n < 1$.  The cosmologies have zero energy density in the initial phase and then the shear viscosity causes the gravitational field to create matter during the evolution. At the final stage, cosmologies are driven to the isotropic Friedmann universe. Although the anisotropy in the universe is smoothed out asymptotically, we point out that there are solutions that simultaneously possess non-positive and non-negative Hubble functions for all time. In view of this fact, we then consider the higher-dimensional theory and show that the cosmological dimensional reduction may work in cosmological models which have shear viscosity. The models extended to higher dimensions have also been analyzed according to the methods described in the present paper, the results show that they all share the same characters. 

   The organization of this paper is as follows. In Sec. II the derivation of two dynamical evolution equations of expansion and shear scalars, which was presented in Ref. [13], is summarized for a convenient reading. In Sec. Ill we consider the axially symmetric Bianchi type I models in which there are only two Hubble functions. The evolutions of the cosmology are described as the flows in the phase space of the Hubble functions. The characteristics of the evolutions 
of the cosmological models for any $n$ are then clarified. The Bianchi type I models with three Hubble functions are discussed in Sec. IV, where we also consider the higher-dimensional models.  Section V is devoted to a summary.  In the Appendix the solutions in the early stage for the $n> 1$ models, which explicitly show how the energy density approaches zero at the initial singularity, are given.

\section {Einsterin Field Equation}

   We consider the $(1 + D)$-dimensional Bianchi type I space-time with the line element 

$$ dr^2 = - dt^2 + \sum_{i=1}^D  a_i^2(t) dx_i^2.    \eqno{(2.1)}$$
\\
The energy-momentum tensor for a fluid with shear viscosity is

$$T_{\mu\nu} = (\epsilon + \bar p) u_\mu u_\nu + \bar p g_{\mu\nu} - \eta \kappa_{\mu\nu},           \eqno{(2.2)}$$
$$\bar p = p- (2/D) \eta u^\mu_{;\mu},                          \eqno{(2.3)}$$
$$\kappa_{\mu\nu} =  u_{\mu; \nu} + u_{\nu; \mu}+ u_\mu n^\lambda u_{\nu;\lambda}  + u_\nu n^\lambda u_{\mu;\lambda} , \eqno{(2.4)}$$
\\
where $\epsilon$  is the energy density, $p$ is the pressure, and $\eta$ is the shear viscosity, respectively. Choosing a comoving frame, where $u^\mu = \delta^0_\mu$, the explicit form of the Einstein equations is

$$R_{\mu\nu}- {1\over 2}g_{\mu\nu} R  = T_{\mu\nu},        \eqno{(2.5)}$$
\\
for the metric (2.1) and the energy-momentum tensor (2.2) can be written as

$$[ (D - 1 )/2D ] W^2 - \sigma^2 = \epsilon,                    \eqno{(2.6)}$$
$${dH_i\over dt} + H_i W - {1\over 2}\left [2{dW\over dt}+(1+{1\over D})W^2 + 2\sigma^2 \right] = p + {2\eta W\over D} - 2\eta H_i, \eqno{(2.7)}$$
\\
where the Hubble functions $H$,, the expansion scalar $W$, and the shear scalar $\sigma^2$ are defined by

 $$H_i ={1\over a_i }{da_i\over dt}, ~~~ W= \sum_i H_i,~~~,2\sigma^2 = \sum_i H_i^2 -{1\over D}W^2. \eqno{(2.8)}$$
\\
The trace part of Eq. (2.5) leads to

$$ 2{dW\over dt } +{1\over D} W^2 + 2\sigma^2 = {2 \over D-1}[\epsilon - D p].      \eqno{(2.9)}$$
\\
Using relation (2.6) to eliminate the $\sigma^2$ in Eq. (2.9) we obtain

$$ {dW\over dt } + W^2 = {D \over D-1} (\epsilon -p).      \eqno{(2.10)}$$
\\
As a consequence of the Bianchi identity we have

$$ {d\epsilon \over dt } + (\epsilon + p)W - 4\eta\sigma^2 = 0.  \eqno{(2.11)}$$
\\
Equation (2.6) can yield a relation

$${d(\sigma^2/W^2)\over dt} = {d(\epsilon/W^2)\over dt}.  \eqno{(2.12)}$$
\\
After substituting the expressions $d\epsilon/dt$ and $dW/dt$ in Eqs.(2.10) and (2.11) into the rhs of Eq. (2.12) we finally obtain the evolutional equation of a shear scalar:

$${d(\sigma^2/W^2)\over dt} = -({\sigma^2\over W^2}) \left [{2D\over D-1} ({\epsilon-p\over W}) + 4 \eta\right]. \eqno{(2.13)}$$
\\
When the universe is filled with stiff matter, i.e., $\epsilon= p$, we can, from Eq. (2.10), find the solution of an expansion scalar:
\\
 $$W= 1/t.                   \eqno{(2.14)}$$
\\
Using relation (2.14), Eq. (2.13) can be written as

$${dy\over y [(D-1)/2D-y]^n} = - 4 \alpha t^{-2n} dt, ~~~ y = {\sigma^2\over W^2}.  \eqno{(2.15)}$$
\\
Equation (2.15) can be integrated exactly if  $n$ is an integer, as has been shown by Banerjee et al. [13].  However, one can only analyze the models for some values of  $n$ as a result of the fact that these integrated relations are too complex. To overcome these difficulties, in Sec. Ill we will analyze the cosmological evolutions by describing them as flows in the phase plane of Hubble functions.

\section{Solution in the Phase Plane}

We first consider the axially symmetric Bianchi type I model in the 1+ 3 dimension; the Hubble functions therein are denoted as

$$H_1=h, ~~~ H_2=H_3 =H,                       \eqno{(3.1)}$$
\\
in terms of which one can, from Eqs. (2.6) and (2.8), obtain

$$W=h+2H, ~~~\sigma^2 ={1\over 2} (h-H)^2, ~~~ \epsilon = H(2h+H).   \eqno{(3.2)}$$
\\
We express $h$ and $H$ in terms of the variables $r$ and $\theta$:

$$h= rsin\theta,~~~  H = rcos\theta.                   \eqno{(3.3)}$$
\\
Thus
$$W =r(sin\theta + 2cos\theta),$$
$$ y={1\over 2} [(1-tan\theta)^2/(2+tan\theta)^2],            \eqno{(3.4)}$$
and Eq. (2.15) can be written as

$$ \int {dy\over y({1\over 2} - y)^n} = -4\alpha \int t^{-2n} dt = {4\alpha\over2n -1} t^{1-2n} + C = {4\alpha\over2n -1} W^{2n-1} + C$$
$$ =  {4\alpha\over2n -1} r^{2n-1} (sin\theta + 2 cos\theta)^{2n-1} + C ,    \eqno{(3.5)}$$
\\
where C is an integration constant. Because $y$  is the function of the variable $\theta $ only, Eq. (3.5) tells us that we can express $r$ as the function of the variable $\theta$.  It is through this property that one can study the cosmological evolutions by analyzing these trajectories in the phase plane. 

{\bf A. Evolutions of the Cosmologies}

{\bf 1. Fixed point}

We only consider the plane where the trajectories are in the regions that satisfy the dominant energy condition [14],  i.e., $\epsilon \geq 0$. The evolutions of the cosmology are therefore confined to the regions $H>0$ and $H + 2h>0$, as Eq. (2.6) shows. The evolutions of the cosmology should start from a fixed point or infinity and then end in another fixed point or infinity in the phase plane.  From Eq. (2.13) we know that the fixed points should be those with zero energy density, which are at $H= 0$ or $H +2h=0$. 

    Furthermore, from Eq. (2.13) we can obtain

$${d\sigma^2/dt\over \sigma^2} = -2W - 4\eta. \eqno{(3.6a)}$$
\\
Since $\eta\geq 0$  and $W= 1/r$, we then obtain a relation

$$\sigma^2 \leq t^{-2}.                                   \eqno{ 3.6b)}$$
\\
Equation (3.6b) tells us that the shear scalar $\sigma^2$ should be a decreasing function and that anisotropy will be smoothed out asymptotically. Therefore, the original point is an attractive fixed point.

{\bf 2. Invariant Lines}

Equation (2.15) can lead to 

$$ {dy\over dt} = -4\alpha W^{2n} y \left({1\over 3} - y \right)^n.  \eqno{(3.7)}$$
\\
Substituting relation (3.4) into Eq. (3.7) we then obtain 

$$ {d\theta\over dt} = {2\alpha\over 3} r^{2n} (cos\theta - sin\theta)(2cos\theta+sin\theta)[(cos\theta+2sin\theta)cos\theta]^n.        \eqno{(3.8)}$$
\\
From the theory of nonlinear differential equations we know that the zeros of Eq. (3.8) give invariant lines in the phase plane of $h \times H$. They are as follows.

(i) The isotropic state is $cos\theta - sin\theta = 0$. In practice, this shall only refer to the original point, resulting from the fact that there is no shear viscosity in the isotropic state. Furthermore, since the shear scalar is a decreasing function, the original point must be the attractive state.

(ii) The states $2cos\theta + sin\theta = 0$ have negative energy density and are thus neglected.

(iii) The states $cos\theta = 0$ or $cos\theta + 2 sin\theta = 0$ have zero energy density.  Because the shear scalar is a decreasing function, the vacuum states must be the initial states.

Using results (i)-(iii) and the exact solution of the $n = 0$ model (note that we can obtain the exact solution for $2n$ = integer models) we therefore conclude that the cosmologies should begin with zero energy density in the initial phase; then the shear viscosity causes the gravitational field to create matter during the evolution; and at the final stage, cosmologies are driven to the isotropic Friedmann universe.

{\bf B. Singularity} 

Equation (3.2) tells us that the energy density can become infinity only if H and/or h are infinite. Equation (2.9) tells us that dW/dt can become infinity only if H and/or h are infinite. Since the Riemann scalar curvature can be expressed as

$$R = 2{dW\over dt} + W^2 + 2H^2 + h^2,  \eqno{(3.9)}$$
\\
one then sees that $R$ can become infinity only if $H$ and/or $h$ are infinite. Therefore, the Einstein initial singularity can arise only if  $H$ and/or $h$ are infinity, i.e., $r \rightarrow \infty$.

    From Eq. (3.5) we know that $r$ can become infinity only if the integrated value on the Ihs is infinity. Since in this section we have discussed that only those states with zero energy density may be the infinite value of $r$, we therefore need only to consider the vacuum states.

   When $y \rightarrow {1\over2}$, then

$$ \int {dy\over y(1/3 -y)^n} \rightarrow 3 \int^{y\rightarrow 1/3} {dy\over (1/3 -y)^n}.    \eqno{(3.10)}$$
\\
The integration (3.10) is infinite only if $n\geq 1$, as easily seen. Thus we have shown that models with $0\leq n< 1$ can avoid the Einstein initial singularity.

{\bf C. Examples}

To give some examples we present the following explicit solutions.

For $n =1/2$,

$$[r(sin\theta +2cos\theta)]^{4\alpha/\sqrt 3} = C {\sqrt{3(1+2tan\theta)} - (2+ tan\theta) \over \sqrt{3(1+2tan\theta)} + (2+ tan\theta)}, \eqno{(3.11)}$$
\\
where $C$ is an integration constant. Using Eq. (3.11) we can plot the trajectories in the phase plane; they then determine the cosmological evolutions. One see that $r$ is finite when $cos\theta = 0$ or $1 + 2 tan\theta = 0$; thus this model has no initial singularity.

For $n = 1$,

$$ ln{(1-2tan\theta)^2\over (1+2tan\theta)} = {4\alpha\over 3 } r (sin\theta + 2cos\theta) + C,  \eqno{(3.12)}$$
\\
where $C$ is an integration constant. Using Eq. (3.12) we can plot the trajectories in the phase plane; they then determine the cosmological evolutions. One see that r is infinite when $cos\theta = 0$ or $1 + 2 tan\theta = 0$; thus this model begins with an initial singularity.

\section{More General Solutions}

We now consider the Bianchi type I models with multiple Hubble functions. From Eq. (2.7) we can obtain 

$${dln(H_i-H_j)\over dt} = {dln(H_i-H_k)\over dt}.  \eqno{(4.1)}$$
\\
Equation (4.1) yields the solutions
$$H_i = (1 - C_i, )H_1 + C_iH_2,~~~ i == 3,....D,         \eqno{(4.2)}$$
\\
where $C$, are the integration constants. Relation (4.2) tells us that one can express all other $D-2$ Hubble functions in terms of only two Hubble functions. Through the same procedures as those described in Sec. Ill we find that the evolutions of the $1 + D$-dimensional Bianchi type I cosmological models containing stiff fluid with shear viscosity as a power function of the energy can also be expressed as flows in the phase plane of $H \times h$... One can also show that the characteristics of the cosmological evolutions are just those described in Sec. Ill, i.e., cosmologies will begin with zero energy density; in the course of evolution the gravitational field will create matter; and finally, cosmologies are driven to the isotropic Friedmann universe. In the same way, we can also prove that there is no Einstein initial singularity in models with $0\leq n < l$.

   To give an illustration we consider the (1 + 3)-dimensional model with three Hubble functions. We now have the relations

$$H_3=(1-C)H_1+ CH_2, ~~~  W=(2-C)H_1 +(1+C) H_2, $$
$$ \sigma^2 = [(1-C+C^2)/3](H_1-H_2)^2.   \eqno{(4.3)}$$
\\
Expressing $H_1$ and $H_2$  in terms of the variables $r$ and $\theta$,

$$H_1 = r cos\theta, ~~~ H_2 = r sin\theta.   \eqno{(4.4)}$$
\\
Then Eq. (2.15) can lead to

$${d\theta \over dt} = {2\alpha\over 3} r^{2n}(1-C+ C^2) (cos\theta-sin\theta)[(2-C)cos\theta+(1+C)sin\theta] [(1-C)cos^2\theta +2sin\theta cos\theta+C sin^2\theta]^n. \eqno{(4.5)}$$
\\
The zeros ofEq. (4.5) give invariant lines in the phase plane. They are as follows.                    

(i) The isotropic state is $cos\theta - sin\theta = 0$. As discussed in Sec. Ill, this shall only refer to the original point, which corresponds to the final state. 

(ii) The states $(2 - C)cos\theta + (1 +C)sin\theta = 0$ can be shown to have negative energy density and are thus nelected.

(iii) The states   $(1 -C)cos^2\theta + 2 cos\theta sin\theta  +C sin^2\theta = 0$ are the vacuum states.   Because shear scalar is a decreasing function, these states must be the initial states.

   Results (i)-(iii) show that the cosmologies will begin with zero energy density and then end in the isotropic Fnedmann universe. One can also use the method described in Sec. Ill to prove that models with $0<n < 1$ have no Einstein
initial singularity. Finally, we want to mention that although the amsotropy is smoothed out asymptotically [as Eq. (3.6b) shows] and cosmologies will be driven to the isotropic Fnedmann universe eventually, there indeed are solutions that possess both nonpositive and non-negative Hubble functions. This
implies that the cosmological dimensional reduction can work in cosmological models which have shear viscosity. This property has been found in our previous paper. Here we show an example in order to complete our discussions of
cosmological models. 

   Consider the $n = 1$ five-dimensional theory with two Hubble functions of  $h$ and $H$ corresponding to those of three-space and extra space, respectively. Using Eqs. (2.14) and (2.15) we obtain

$$ W = t^{-1} , ~~~ \sigma^2 = {3\over 8} (1/t^2)[1+C e^{-3\alpha /2t}]^{-1}.   \eqno{(4.6)}$$
\\
where $C$ is a positive constant. With the definitions of $W$ and $\sigma^2$ in Eq. (2.8) we then obtain 

$$ h = (1/4t)(1 + [1 + C exp( - 3\alpha /2t)]^{ -l/2}),  $$
$$ H = (1/4t)(1 -3 [1 + C exp( - 3\alpha /2t)]^{ -l/2}),  \eqno{(4.7)}$$
$$ h = (1/4t)(1 - [1 + C exp( - 3\alpha /2t)]^{ -l/2}),  $$
$$ H = (1/4t)(1 +3 [1 + C exp( - 3\alpha /2t)]^{ -l/2}),  \eqno{(4.8)}$$
\\
Solution (4.7), in which $h$ is positive while $H$ is negative for all time, can be found when $C < 8$. We have also checked that the simultaneous existence of
non-positive and non-negative Hubble functions is found in other (including $1+3$ 3) space-time models with other values of $n$.  Thus we have clarified the effects of shear viscosity on the characteristics of cosmological evolution.

\section{Conclusions}

In this paper we have discussed Bianchi type I cosmological models with a viscous fluid, assuming that the shear viscosity is a power function of the energy density, such as $\eta =\alpha \epsilon^n$ We have presented a detailed study of these models by describing the evolutions of cosmologies as the flows  the phase plane of Hubble functions. We have clarified the property of these models with any value of $n$.  In  particular We have proved that there are no Einstein initial singulanties in models with $0<n <1$. The cosmologies have been found to begin with zero energy density; then the shear viscosiy, causes the gravitational field to create matter during the evolution. At the final stage, cosmologies are driven to the isotropic Friedmann universe. We have also pointed out that there are solutions that possess both non-positive and non-negative Hubble functions for all time. In view of this fact, we then considered five-dimensional theory and gave a solution that explicitly showed that the cosmological dimensional reduction can work in cosmological models which have shear viscosity.  Models extended to other dimensions can also be analyzed according to the procedures described in this paper; the results show that they all share the same characteristics.

     The models considered in this paper are only concerned with stiff matter. The same models with other matter fields are certainly of interest and remain to be studied.

\section{Appendix: Solution in the Early Stage}

   Since the behavior of the vanishing energy density at the initial singularity, which is argued in Sec. Ill to be a common property belonging to $n \geq 1$ models, seems unusual, we now give the solution in the early stage to explicitly show how the energy density approaches zero asymptotically.

 Let $\epsilon = p$ and $D=3$ in Eqs. (2.7) and (2.9). We obtain

$${dH_i\over dt} +H_i W = {2\over 3} \eta W -2 \eta H_i,   \eqno{(A1)}$$
\\
From the discussions given in Sec. Ill we know that $n> 1$ models will begin along the invariant lines $H=0$ or $2h + H = 0$. Thus from Eq. (Al) we find the approximate solution as $H\rightarrow 0$. (The case of $2h +H \rightarrow 0$ could be analyzed in the same way and will give the same conclusion)

     To be consistent with Eq. (2.14), $W=h+2H= t^{-1}$ , one can define

$$ h=t^{-1} - \delta h,                          \eqno{(A2a)}$$
$$H=\delta h/2,                               \eqno{(A2b)}$$
\\
where $0 \leq \delta h \leq 1$. Using the relation of Eq. (3.2), $\epsilon = H(2h+H)$, implies

$$\epsilon \approx t^{-1} \delta h.  \eqno{(A3)}$$
\\
Let $\eta = \alpha \epsilon^n$; we then approximate Eq. (Al) as 

$${\delta h \over dt} + t^{-1} \approx {4\alpha\over 3} t^{-1}({\delta h \over dt})^n.        \eqno{(A4)}$$
\\
The solution for the $n = 1$ model is

$$\delta h \approx  C t^{-1}exp(-4\alpha/3t),                \eqno{(A5a)}$$
$$\epsilon \approx  C t^{-2}exp(-4\alpha/3t),                \eqno{(A5b)}$$
\\
where$C$ is an integration constant. Solution (A5) is consistent with the asymptotic form of the exact solution found in Eq. (3.9) of Ref. 13. The solution for the $n > 1$ model is 

$$\delta h \approx  [4\alpha(n-1)/3(2n-1)]^{1/(1-n)}   t^{n/(n-1)}, \eqno{(A6a)}$$
$$\epsilon \approx  [4\alpha(n-1)/3(2n-1)]^{1/(1-n)}   t^{1/(n-1)}, \eqno{(A6a)}$$
\\
Solutions (A5) and (A6) explicitly show that $n geq 1$ models have vanishing energy density at the initial singularity.

%%%%%%%%%%%%%%%%%%%%%%%
\newpage
\begin{enumerate}
\item C M. Misner, Nature 214, 40 (1967); Astrophys. J. 151, 431 (1967).
\item S. Weinberg, Astrophys. J. 168, i?5 (1972).
\item S. Wcinberg, Gravitation and Cosmology (Wiley, New York, 1972).
\item P. Langacker. Phys. Rep. 72. 185 (1981).
\item  I. Waea. R. C. Falcao, and R. Chanda, Phys. Rev. D 33. 1839 (1986).
\item T. Pacher, J. A. Stein-Schabes. and M. S. Turner, Phys. Rev. D 36, 1603
(1987).
\item A. H. Guth, Phys. Rev. D 23, 347 (1981).
\item G. L. Murphy, Phys. Rev. D 8, 4231 (1973).
\item N. 0. Sanros, R. S. Dias, and A. Banerjee, J. Math. Phys. 26. 878 (1985). 
\item W.-H. Huang, Phys. Lett A 129, 429 (1988).
\item W.-H. Huang, J. Math. Phys., J. Math. Phys., Vol. 31, 1456 (1990). [gr-qc/0308059]
\item V. A. Belinsku and V. M. Khalatnikov, Sov. Phys. JETP 42, 205 (1976);
Sov. Phys. JETP 45, 1 (1977).
\item A. Banerjee, S. B. Duttachoudhury, and A. K. Sanyal, J. Math. Phys. 26,
3010 (1985).
\item S. W. Hawking and G. F. R. Ellis, The Large Scale Structure of'Spacetime
(Cambridge U.P., Cambridge, 1973).
\item W.-H. Huang, Phys. Lett. A 136, 21 (1989).

\end{enumerate}

\end{document}